# On-chip integrated Yb$^{3+}$-doped waveguide amplifiers on thin film lithium niobate

ZHIHAO ZHANG,[1,5,6] ZHIWEI FANG,[2,10] JUNXIA ZHOU,[2,3] YOUTING LIANG,[2,3] YUAN ZHOU,[1,5] ZHE WANG,[1,5,6] JIAN LIU,[2,3] TING HUANG,[2,4] RUI BAO,[2,3] JIANPING YU,[1,5] HAISU ZHANG,[2] MIN WANG,[2,8] AND YA CHENG[1,2,3,7,8,9,11]

[1]*State Key Laboratory of High Field Laser Physics and CAS Center for Excellence in Ultra-intense Laser Science, Shanghai Institute of Optics and Fine Mechanics (SIOM), Chinese Academy of Sciences (CAS), Shanghai 201800, China*
[2]*The Extreme Optoelectromechanics Laboratory (XXL), School of Physics and Electronic Science, East China Normal University, Shanghai 200241, China*
[3]*State Key Laboratory of Precision Spectroscopy, East China Normal University, Shanghai 200062, China*
[4]*School of Physics and Electronic Science, East China Normal University, Shanghai 200062, China*
[5]*Center of Materials Science and Optoelectronics Engineering, University of Chinese Academy of Sciences, Beijing 100049, China*
[6]*School of Physical Science and Technology, ShanghaiTech University, Shanghai 200031, China*
[7]*Collaborative Innovation Center of Extreme Optics, Shanxi University, Taiyuan 030006, China*
[8]*Collaborative Innovation Center of Light Manipulations and Applications, Shandong Normal University, Jinan 250358, People's Republic of China*
[9]*Shanghai Research Center for Quantum Sciences, Shanghai 201315, China*
[10]*zwfang@phy.ecnu.edu.cn*
[11]*ya.cheng@siom.ac.cn*

**We report the fabrication and optical characterization of Yb$^{3+}$-doped waveguide amplifiers (YDWA) on the thin film lithium niobate fabricated by photolithography assisted chemo-mechanical etching. The fabricated Yb$^{3+}$-doped lithium niobate waveguides demonstrates low propagation loss of 0.13 dB/cm at 1030 nm and 0.1 dB/cm at 1060 nm. The internal net gain of 5 dB at 1030 nm and 8 dB at 1060 nm are measured on a 4.0 cm long waveguide pumped by 976nm laser diodes, indicating the gain per unit length of 1.25 dB/cm at 1030 nm and 2 dB/cm at 1060 nm, respectively. The integrated Yb$^{3+}$-doped lithium niobate waveguide amplifiers will benefit the development of a powerful gain platform and are expected to contribute to the high-density integration of thin film lithium niobate based photonic chip.**

Featured with high-efficiency and broad-gain bandwidth, the ytterbium ion (Yb$^{3+}$) doped laser and fiber amplifiers operating near the 1 micrometer (μm) wavelength have been extraordinarily successful in industrial processing of materials, as such devices can be operated with extremely high output power [1, 2]. In addition to bulk solids and fiber, the Yb$^{3+}$-doped integrated waveguide lasers and amplifiers have become a new technology for the on-chip applications. The on-chip Yb$^{3+}$-doped waveguide lasers and amplifiers have been previously demonstrated on various substrates including glasses, Tantalum pentoxide (Ta$_2$O$_5$), Aluminum oxide (Al$_2$O$_3$), Yttrium-Aluminium-Granat (YAG), Calcium fluoride (CaF$_2$), Potassium yttrium tungstate (KY(WO$_4$)$_2$), and Yttrium lithium fluoride (LiYF$_4$) [3-14]. However, these material platforms cannot simultaneously support ultralow propagation loss, highly-efficient optical nonlinearities and ultrafast electro-optical modulation. Lithium niobate (LN) provides an attractive option to be alternative host material for rare earth ions, besides their broad optical transparency window, low optical loss, high refractive index, high nonlinear coefficient, and large electro-optical effect. The on-chip Yb$^{3+}$-doped waveguide have already been demonstrated on weakly guiding ion-diffused waveguides in Yb$^{3+}$-doped bulk LN [15-17]. These waveguides feature weak optical confinement and large bending radius due to the low index contrast. Therefore, they are unsuitable for dense integration and also challenging to realize high-gain waveguide amplifiers. In the last two decades, the rapid developments of the wafer-scale, high-quality, thin film lithium niobate on insulator (TFLNOI) and micro- and nano-fabrication techniques have realized on-chip integrated TFLNOI photonic devices with unprecedented performance [18, 19]. The on-chip microlasers and amplifiers based on the Er$^{3+}$-doped TFLNOI have been demonstrated recently, showing great promise for high-performance scalable light sources based on integrated photonics [20-27]. The on-chip Yb$^{3+}$-doped microlasers based on microresonator have also been demonstrated on the TFLNOI recently [28, 29]. Yet to date the on-chip Yb³⁺-doped amplifiers has not been demonstrated on TFLNOI. Here, we demonstrate for the first time the on-chip Yb³⁺-doped lithium niobate waveguide amplifiers fabricated by photolithography-assisted chemo-mechanical etching. The fabricated Yb³⁺-doped lithium niobate waveguides demonstrates low propagation loss of 0.13 dB/cm at 1030 nm and 0.1 dB/cm at 1060 nm based on the cut-back measurements of different lengths of waveguides. The internal net gain of 5 dB at 1030 nm and 8 dB at 1060 nm are demonstrated on a 4.0 cm long waveguide pumped by 976nm laser diodes which indicates the gain per unit length are 1.25 dB/cm at 1030 nm and 2 dB/cm at 1060 nm. The integrated Yb³⁺-doped lithium niobate waveguide amplifiers will benefit the development of a powerful gain platform and are expected to contribute to the high-density integration of TFLNOI-based photonic chip.

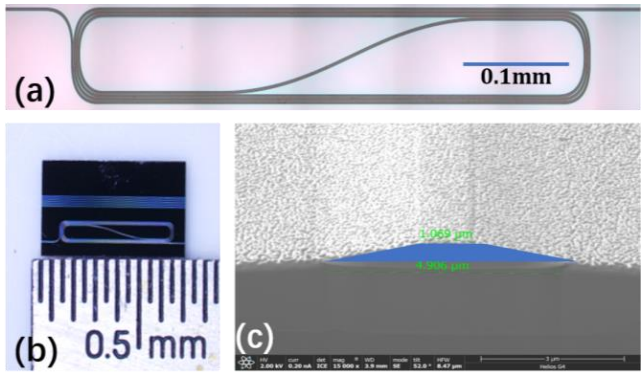

Fig. 1. (a) Optical micrograph of the fabricated Yb³⁺-doped spiral waveguide. (b) The photograph of the Yb³⁺-doped TFLNOI chip with different lengths of waveguides. (c) The false color scanning electron microscope (SEM) images of the cross section of the fabricated Yb³⁺-doped waveguide (blue).

The on-chip integrated Yb³⁺-doped LN waveguides is fabricated on a 600-nm-thickness Z-cut TFLNOI with Yb³⁺ concentration of 0.5 mol% using photolithography-assisted chemo-mechanical etching, and the more details about the LN waveguides fabrication can be found in our previous work [30]. As shown in Figure 1(a), the fabricated Yb³⁺-doped waveguide with the length of 4 cm is folded in a spiral layout for dense integration. Figure 1(b) shows the optical photograph of the Yb³⁺-doped TFLNOI chip with different lengths of waveguides. Figure 1(c) shows the scanning electron microscope (SEM) image of the cross section of the fabricated Yb³⁺-doped LN waveguide with a top-width of ~1.1 μm and a bottom-width of ~4.9 μm. Figure 2(a) shows the energy level diagram of Yb³⁺, which is the basis for achieving the amplification at different signal light wavelengths. Figures 2 (b) and (c) showing situ quantitative elements analysis of the surface of fabricated Yb³⁺-doped waveguide measured by energy dispersive spectrometer (EDS) (Zeiss Gemini SEM450). As shown in figure 2(b), ytterbium element is detected. The map analysis by EDS in figure 2(c) corroborates the uniform micro-scale distribution of ytterbium element in waveguide samples. This provides a favorable factor for the waveguide amplifier to achieve uniform amplification over the entire transmission length.

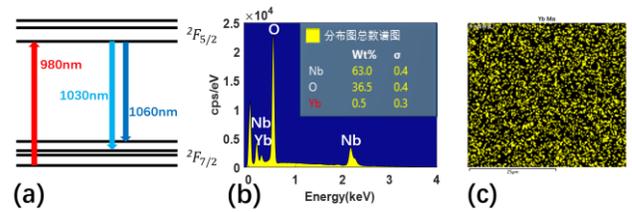

Fig. 2. (a) The energy level diagram of Yb³⁺. (b) The in situ quantitative elements analysis of the surface of the fabricated YDWA measured by energy dispersive spectrometer (EDS). (c) The map of ytterbium element analysis in waveguide samples by of EDS.

The experimental setup to characterize optical amplification performance of YDWA is illustrated in Fig. 3(a). Here, a pump laser at 976 nm is provided by a laser diode (CM97-1000-76PM, II-VI Laser Inc.), while laser diodes at wavelength 1030 nm and 1060 nm (II-VI Laser Inc.) are used as the signal light. The polarization states of both the pump and signal laser beams are adjusted using in-line fiber polarization controllers (FPC561, Thorlabs Inc) to achieve the best amplification performance. The pump and signal laser beams are combined and seperated by the fiber-based wavelength division multiplexers (WDM) at the input port of the on-chip integrated Yb³⁺-doped waveguide with lensed fibers. The insets of Figure 3(a) display the the Yb³⁺-doped waveguide butt-coupled by lensed fibers. Both the output light signals are analyzed by an spectrometer (NOVA, Shanghai Ideaoptics Corp., Ltd). The powers of the input and output pump and signal laser are measured by a power meter (PM100D, Thorlabs Inc.). The output of the waveguide was also zoomed in imaging by an objective onto an infrared camera (InGaAs Camera C12741-03, Hamamatsu Photonics Co., Ltd.) for the observation output mode profiles. As shown in figures 3(b),(c) and (d), both the pump laser and signal laser are in the fundamental mode of the Yb³⁺-doped waveguide.

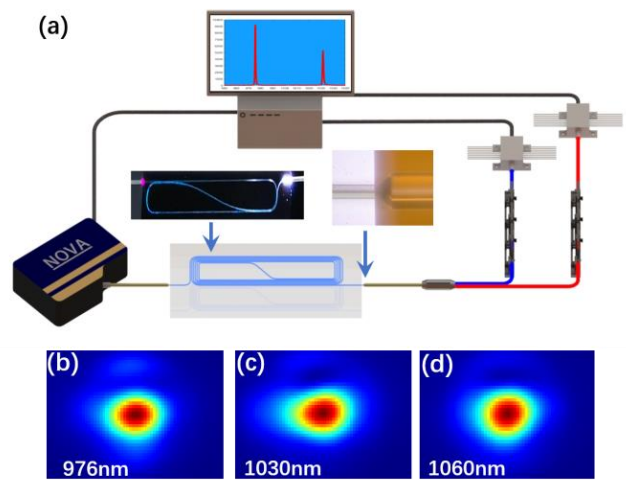

Fig. 3 (a) The experimental setup to characterize optical amplification of YDWA. Insets show the photograph of the YDWA (left) and the coupling lensed fiber (right). The infrared images of the output port of the fabricated Yb³⁺-doped waveguide with wavelength (b) 976 nm laser beam, (c) 1030 nm laser beam and (d) 1060 nm laser beam.

Before characterizing the gain of the Yb³⁺-doped LN amplifier, the loss of the waveguide amplifier needs to be determined first. We use the cut-back method to measure the loss of the pump light and the signal light in the waveguide. The propagation loss of the waveguide as $\alpha_l$ includes the absorption loss induced by ground-state $Yb^{3+}$ ions and the scattering loss induced by the surface roughness of waveguide. Setting $\alpha_0$ as the fiber-to-chip coupling losses per facet, and $L$ as the length of $Yb^{3+}$-doped LN waveguide amplifier, the transmittance of the waveguide can be expressed as $T(L) = -\alpha_l L - 2\alpha_0$. The waveguide transmittance is the ratio of the power of the pump light (signal light) input into the lens fiber over the output power from the lens fiber. The loss curves shown in Figure 4 were fitted by measuring the transmittance of waveguides with different lengths of 0.8 cm, 1 cm, 2 cm and 4 cm. Figures 4(a), (b) and (c) show the loss curves of 976nm pump light, 1030nm signal light and 1060nm signal light respectively. Through linear fitting, it can be deduced that the propagation loss of the pump light at 976 nm is 6.78 dB/cm, while the propagation loss of the signal light are 0.13 dB/cm at 1030 nm and 0.1 dB/cm at 1060 nm respectively. Obviously, the transmission loss depends on different wavelength mainly because $Yb^{3+}$ ions have different absorption coefficient for different wavelength [31]. The coupling losses of the pump light is 20 dB per facet, while the signal light is 13 dB per facet and 9 dB per facet, respectively. The main reason for the high coupling loss is due to the mismatch between the spot size of the waveguide and the lensed fiber (~2 μm diameter).

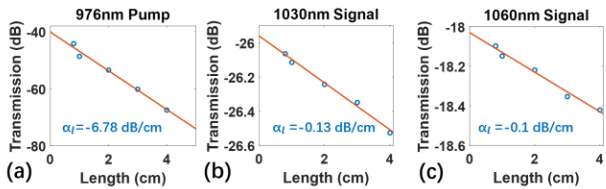

Fig. 4. The measured loss curve of the pump light at a)976 nm and the signal light at (b)1030 nm and (c)1060 nm.

Then we measured the internal net gain in a 4.0 cm long $Yb^{3+}$-doped LN waveguide amplifier. The internal net gain is measured by the signal-enhancement method. It was defined by the following equation:

$$g = 10\lg\frac{P_{on}}{P_{off}} - \alpha_l L$$

where $P_{on}$ and $P_{off}$ are the output powers of the signal light with and without the excitation of the pump light, respectively, and $\alpha_l$ is the optical propagation loss per unit length. Figures 5(a) and (b) present the measured signal spectra at 1030 nm and 1060 nm with the increasing pump powers, which apparently shows the signal enhancement. Figure 5(c) shows the net gain as a function of the increased pump power with fixed signal powers at 1030 nm. The waveguide amplifier optical gain increases rapidly at the small pump powers, and tends to saturate at relatively high pump powers (>13 mW). The maximum internal net gain of ~ 5 dB is achieved. Subsequently, the wavelength of the signal light was changed to 1060 nm, the net gain curve was shown in Figure 5(d). A similar signal amplification phenomenon can be seen, the maximum internal net gain of ~ 8 dB is achieved with the pump power at 48 mW, which is higher than 1030 nm.

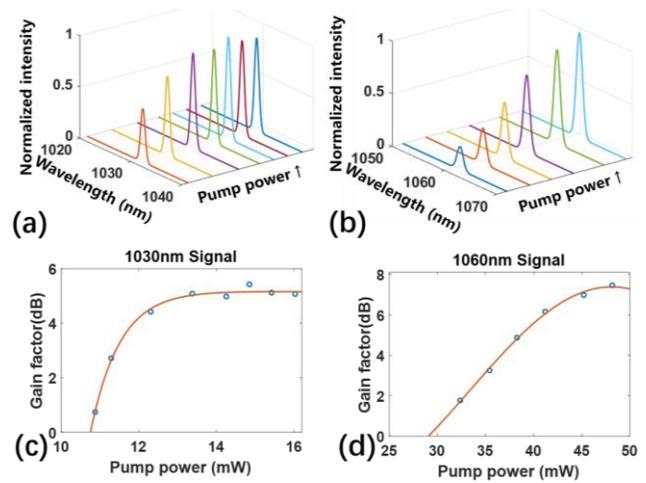

Fig. 5 Gain characterization of the $Yb^{3+}$-doped LN waveguides. Measured signal spectra as a function of increasing pump powers measured at (a)1030 nm and (b)1060 nm. Measured net internal gain as a function of increasing pump powers at different signal wavelength (c)1030 nm and (d)1060nm.

As shown in Figure 6(a) and (b), when the signal light gain tends to saturated, a significant second harmonic generation (~488 nm) from the pump light at 976 nm was appeared. It can be seen with the naked eye on the surface of waveguide amplifier, as shown in the right inset in Figure 3(a). And the peak at 508 nm and 501 nm correspond to the sum frequency generations from the pump light and signal light at 1060 nm and 1030 nm, respectively.

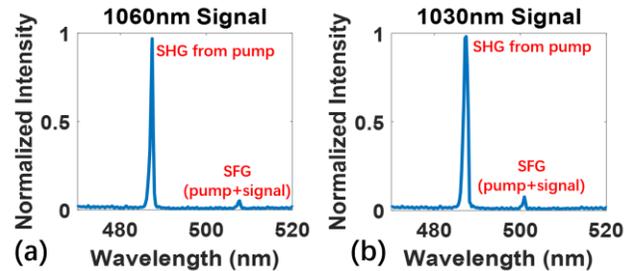

Fig. 6. The spectral details of nonlinear frequency conversion from the signal light at (a)1060 nm and (b)1030 nm and the pump light at 976 nm.

In conclusion, we have demonstrated the fabrication and optical characterization of YDWA on the thin film lithium niobate fabricated by photolithography assisted chemo-mechanical etching. The fabricated $Yb^{3+}$-doped lithium niobate waveguides demonstrates low propagation loss of 0.13 dB/cm at 1030 nm and 0.1 dB/cm at 1060 nm. The internal net gain of 5 dB at 1030 nm and 8 dB at 1060 nm are demonstrated on a 4.0 cm long waveguide pumped by 976nm laser diodes which indicates the gain per unit length are 1.25 dB/cm at 1030 nm and 2 dB/cm at 1060 nm. Further optimizations concerning the $Yb^{3+}$ ions doping concentration to increase the absorption for pump light, and the waveguide geometric design to allow for high-power amplifications, are anticipated to increase the waveguide gain to even higher values, showing the potential of the on-chip applications for photonic integrated circuit (PIC).